\documentclass[floatfix,aps,twocolumn,showpacs,superscriptaddress,pra,10pt,longbibliography]{revtex4-2}
\usepackage{amsmath,amsfonts,amssymb,graphicx}
\usepackage{epstopdf}
\usepackage{bm}
\usepackage{hyperref}

\usepackage[utf8]{inputenc}
\usepackage[T1]{fontenc}
\usepackage[english]{babel}

\IfFileExists{newtxtext.sty}
    {\usepackage{newtxtext,newtxmath}}
    {\IfFileExists{stix.sty}
       {\usepackage{stix}}
       {\IfFileExists{mathptmx.sty}
       {\usepackage{mathptmx}}{} } }

\usepackage{hyperref}
\usepackage{xcolor}

\begin{document}
\title{
Quantum Monte Carlo study of the role of p-wave interactions in ultracold repulsive Fermi gases}

\author{Gianluca Bertaina}
\affiliation{Istituto Nazionale di Ricerca Metrologica, Strada delle Cacce 91, I-10135 Torino, Italy}
\author{Marco G. Tarallo}
\affiliation{Istituto Nazionale di Ricerca Metrologica, Strada delle Cacce 91, I-10135 Torino, Italy}
\author{Sebastiano Pilati}
\affiliation{School of Science and Technology, Physics Division, Universit{\`a} di Camerino, 62032 Camerino (MC), Italy}
\affiliation{INFN, Sezione di Perugia, I-06123 Perugia, Italy}

\begin{abstract}
Single-component ultracold atomic Fermi gases are usually described using noninteracting many-fermion models. However, recent experiments reached a regime where $p$-wave interactions among identical fermionic atoms are important. 
In this paper, we employ variational and fixed-node diffusion Monte Carlo simulations to investigate the ground-state properties of single-component Fermi gases with short-range repulsive interactions. 
We determine the zero-temperature equation of state, and elucidate the roles played by the $p$-wave scattering volume and the $p$-wave effective range.
A comparison against recently-derived second-order perturbative results shows good agreement in a broad range of interaction strength.
We also compute the quasiparticle effective mass, and we confirm the perturbative prediction of a linear contribution in the $p$-wave scattering volume, while we find significant deviations from the beyond-mean-field perturbative result, already for moderate interaction strengths.
Finally, we determine ground-state energies for two-component unpolarized Fermi gases with both interspecies and intraspecies hard-sphere interactions, finding remarkable agreement with a recently derived fourth-order expansion that includes $p$-wave contributions.
\end{abstract}

\maketitle

\section{Introduction}
In ultracold atom theory, single-component (i.e., fully spin-polarized) Fermi gases are usually treated as noninteracting systems~\cite{Giorgini2008}. The reason is that $s$-wave scattering, which dominates at low temperature and density, is inhibited by the Pauli exclusion principle, while $p$-wave scattering is strongly suppressed at ultracold temperatures, as observed in several seminal experiments~\cite{Gupta2003,Zwierlein2003,Omran2015}. 
Previously, various other experiments employed Feshbach resonances to enhance the role of $p$-wave interactions (see, e.g., Refs.~\cite{Jin2003,Salomon2004,Jin2008,Mukaiyama2008,Mukaiyama2018}), but they were affected by strong losses due to inelastic collisions. However, recent experiments reached a high-density regime where $p$-wave interactions between fermionic atoms~\cite{Top_Spinpolarizedfermionswave_2021,Venu_Unitarypwaveinteractions_2023} or molecules~\cite{Duda_LonglivedfermionicFeshbach_2022} play a relevant role. 
Spin-polarized Fermi gases are also employed as the initial state in high-precision atomic clocks~\cite{LudlowRMP2015}. For this application, the interaction energy due to $p$-wave intraspecies scattering must be precisely known, since it induces a shift in the clock frequency~\cite{Lemke_WaveColdCollisions_2011,Ludlow_Coldcollisionshiftcancellationinelastic_2011,Bishof_Inelasticcollisionsdensitydependent_2011,Martin_QuantumManyBodySpin_2013}.
Furthermore, experiments performed with single-component Fermi gases have recently allowed observing Pauli blocking of light scattering~\cite{DebPauliBlocking,Margalit_Pauliblockinglight_2021}.

So far, most theoretical and computational studies addressed interaction effects in two-component Fermi gases, focusing on $s$-wave interspecies interactions.
In particular, the zero-temperature equation of state (EOS) of spin-balanced systems has been determined using second-order perturbation theory in the seminal article by Lee and Yang~\cite{LeeYang} in 1957. This result has been extended to the third-order term~\cite{DeDominicis,Efimov1966135,Bishop1973,Kaiser2011} and, recently, to the fourth-order term~\cite{Wellenhofer_Effectivefieldtheory_2021} in the $s$-wave scattering length $a_s$ (see also previous estimates~\cite{Baker_Neutronmattermodel_1999}). 
Importantly, these expansions have been compared against nonperturbative results, in particular against quantum Monte Carlo simulations~\cite{Conduit_InhomogeneousPhaseFormation_2009,2010Pilati,Trivedi,pilati2014}, which provide rigorous upper bounds to the ground-state energies of fermionic systems.
These comparisons turned out to be fruitful, allowing one to shed light on the regimes of validity of various quantum many-body techniques.
In contrast, Fermi gases with imbalanced populations are more poorly understood.
Perturbation theories did address also spin-imbalanced gases~\cite{Kanno1,Kanno2,PhysRevLett.95.230403,PhysRevA.90.023605,Chankowski_Groundstateenergypolarized_2021,https://doi.org/10.48550/arxiv.2206.06932,Pera_Itinerantferromagnetismdilute_2023,https://doi.org/10.48550/arxiv.2206.05076}, but intraspecies interactions were not included, or they were described only up to the first order in the $p$-wave scattering volume $v$ and without including effects due to the $p$-wave effective range $R$.
For the single-component (i.e., fully polarized) Fermi gas, the ground-state properties have only recently been computed up to the second order in $v$~\cite{Ding_FermiLiquidDescriptionSingleComponent_2019}. It was pointed out that, already at this order, the role of the $p$-wave effective range $R$ must be accounted for~\cite{Ding_FermiLiquidDescriptionSingleComponent_2019,Maki_RoleEffectiveRange_2020,Maki_TransportwaveinteractingFermi_2023}.
To the best of our knowledge, quantum Monte Carlo (QMC) results for dilute single-component Fermi gases with intraspecies interactions have not been provided yet.

In this paper, we employ variational and fixed-node diffusion Monte Carlo (DMC) simulations to determine the ground-state properties of single-component atomic Fermi gases with short-range repulsive interactions. 
In particular, the ground-state energy and the quasiparticle effective mass are determined, exploring different regimes of interaction strength.
Two models for the interatomic potentials are considered, namely, the hard-sphere (HS) and the soft-sphere (SS) potentials. This allows us to separately analyze the roles played by $v$ and by $R$.
Our QMC results are compared against the beyond-mean-field (BMF) expansion of Ref.~\cite{Ding_FermiLiquidDescriptionSingleComponent_2019}. In the case of the ground-state energy, good agreement is found in a broad range of the interaction parameters.
For the effective mass, we confirm the unusual perturbative prediction of a dominant linear term in the $p$-wave scattering volume in a narrow interaction regime. However, we observe significant discrepancies in beyond-mean-field corrections, possibly indicating that an accurate determination of this Fermi-liquid parameter requires higher-order terms, or that the variational nodal surface should be improved.
For completeness, we also determine the ground-state energy of a two-component Fermi gas with both interspecies and intraspecies (hard-sphere) interactions, in contrast to various previous studies that considered only interspecies interactions (intraspecies interactions have been addressed in Ref.~\cite{AriasdeSaavedra_Ferromagnetictransitiontwocomponent_2012}, but only within the Fermi hypernetted-chain theory).
A comparison with the recently derived fourth-order expansion~\cite{Wellenhofer_Effectivefieldtheory_2021} shows very good agreement. This represents an important cross-validation between perturbative results and variational upper-bounds provided by QMC simulations.

The paper is organized as follows:
the model Hamiltonian and the scattering parameters are defined in Section~\ref{sec:models}.
The QMC methods to determine the expectation values are briefly described in Section~\ref{sec:methods}. The two procedures used to estimate the quasi-particle effective mass are reviewed in more detail in Subsections~\ref{subsec:dispersion} and~\ref{subsec:finitemass}.
Our results for the ground-state energy and the effective mass of fully polarized gases are presented in Section~\ref{sec:polarized}, while those for balanced repulsive gases are discussed in Section~\ref{sec:balanced}.
In Section~\ref{sec:conclusions} we draw our conclusions and provide some perspective for future research.
In Appendix~\ref{app:scattering}, we report further details on partial-wave scattering for the considered short-range potentials.

\section{Model}\label{sec:models}
We consider a nonrelativistic ensemble of $N$ (pseudo-)spin $1/2$ fermions in three dimensions, described by the following Hamiltonian in continuous space:
\begin{equation}
    H = -\frac{\hbar^2}{2m}\sum_{\substack{\sigma=\uparrow,\downarrow\\i=1}}^{N_\sigma}\nabla_{\sigma,i}^2 + \sum_{\substack{\sigma=\uparrow,\downarrow\\i<j}}^{N_\sigma}V_{\sigma\sigma}(r_{ij}) +\sum_{i,i'}^{N_\uparrow, N_\downarrow}V_{\uparrow\downarrow}(r_{ii'}) \,,
\end{equation}
where $m$ is the mass, ${\bf r}_i$ represents the three-dimensional (3D) coordinates of the $i$-th particle, $r_{ij}=|{\bf r}_i-{\bf r}_j|$ is the distance between the $i$-th and the $j$-th particles, and we indicate the two relevant internal states by $\sigma=\uparrow,\downarrow$. These can refer to hyperfine states, in the case of alkali-metal atoms or the ground-state manifold of alkaline-earth atoms, or orbital states, when considering the clock states of alkaline-earth atoms. We focus on short-range intraspecies ($V_{\uparrow\uparrow}$ and $V_{\downarrow\downarrow}$) and interspecies ($V_{\uparrow\downarrow}$) interaction potentials. 

Since we include neither spin-spin nor spin-orbit interactions, nor spin-flipping external fields, the total spin populations $N_\uparrow$ and $N_\downarrow$ are separately conserved, besides their sum $N$. Simulations are performed with periodic boundary conditions (PBC) in a cubic box of size $L=(N/n)^{1/3}$, where $n=n_\uparrow+n_\downarrow$ is the total particle density and $n_{\sigma}$ are the partial densities. The Fermi wavevector of the $\uparrow$ component is $k_F=(6\pi^2 n_\uparrow)^{1/3}$ and its Fermi energy is $E_F=\hbar^2 k_F^2/2m$.

In the low-energy limit, the above Hamiltonian is customarily simplified by setting $V_{\sigma\sigma}=0$, since the scattering amplitude for two fermions with the same spin does not contain even-wave contributions, in particular the $s$-wave term, while higher-order terms, starting from the $p$-wave one, are suppressed, so that the interspecies interactions quantitatively dominate the EOS. However, in the case of spin-polarized systems, or when accurate determination of the EOS is required, taking into account at least the $p$-wave contribution from $V_{\sigma\sigma}$ is crucial.

Close to broad $s$-wave Feshbach resonances, the $s$-wave scattering length $a_s$ is sufficient to describe the effects of short-range interactions in a fermionic atomic gas~\cite{Simonucci_Broadvsnarrow_2005}. Conversely, it has been argued~\cite{Luciuk_Evidenceuniversalrelations_2016,Ding_FermiLiquidDescriptionSingleComponent_2019,Maki_RoleEffectiveRange_2020,Maki_TransportwaveinteractingFermi_2023} that an accurate description of $p$-wave dominated Fermi gases requires specification of both the $p$-wave scattering volume $v$ and the $p$-wave effective range $R$. 
Although the consideration of a van der Waals tail may be relevant for $p$-wave effects of atomic potentials (see discussion in the Conclusions~\ref{sec:conclusions}), here we focus on the strictly short-range case, to compare with existing literature~\cite{Ding_FermiLiquidDescriptionSingleComponent_2019}. In order to have analytical expressions for the scattering parameters, we consider SS potentials $V_{S}(r)= V_0$ for $r \le R_{S}$, and $V_{S}(r)= 0$ for $r > R_{S}$, using various combinations of the strength $V_0>0$ and the diameter $R_S$ so as to independently vary $v$ and $R$. The HS limit corresponds to taking $V_0\to\infty$. See App.~\ref{app:scattering} for a recap of scattering theory for these potentials.

In this paper, we mostly focus on the fully polarized case with $N=N_\uparrow$ and $N_\downarrow=0$, where $p$-wave collisions give the dominant contribution to the interaction energy. However, for more generality we consider also the spin-balanced case with $N_\downarrow=N_\uparrow$, where $p$-wave collisions contribute only to subleading order in the interaction strength. In the fully polarized case (Sec.~\ref{sec:polarized}), we compare the results when setting $V_{\uparrow\uparrow}$ either equal to a SS potential with $v=R_S^3/24$ (SS24), or with $v=R_S^3/12$ (SS12), or to the HS potential, which corresponds to $v=R_S^3/3$. These three choices amount to fixing three values for the parameter $K_0 = (m V_{0})^{1/2} R_S/\hbar$, which relates the strength and range of the potentials. The weak repulsion SS24 potential is close to a SS model in which the $s$-wave effective range is null, while the SS12 potential is representative of an intermediate regime (see Fig.~\ref{fig:scatteringSS}). In the spin-balanced case (Sec.~\ref{sec:balanced}), we compare the case of a HS potential present only in the opposite-spin sector $V_{\uparrow\downarrow}$, with $V_{\downarrow\downarrow}=V_{\uparrow\uparrow}=0$ and $a_s=R_s$, to the case in which the same potential is also present in the equal-spin sector $V_{\downarrow\downarrow}=V_{\uparrow\uparrow}=V_{\uparrow\downarrow}$.

\section{Methods}\label{sec:methods}
The QMC methods that we employ are the variational Monte Carlo (VMC) and fixed-node DMC methods, which stochastically solve the many-body Schr\"odinger equation, either with a variational wavefunction $\Psi_T$~\cite{CeperleyMonteCarlosimulation1977}, or with an imaginary-time-projected wavefunction $\Psi_{\tau}=e^{-\tau \hat{H}}\Psi_T$, whose nodal surface (where the wavefunction is zero) is constrained to that of the guiding function $\Psi_T$ so as to remove the fermionic sign problem~\cite{Reynolds_FixednodequantumMonte_1982}. In both cases expectation values of various operators can be obtained: in particular, both methods provide an upper bound for the exact ground-state energy. While the precision of the results can be increased at will, by reducing the errorbar due to stochastic sampling, the accuracy of the results is affected by a few factors. For VMC, results fully depend on the chosen $\Psi_T$, which must therefore be suitably optimized; for DMC, results are affected only by inaccuracies in the nodal surface, besides finite time-step and walker-population-size biases, which can be properly extrapolated to zero~\cite{DePasquale_Reliablediffusionquantum_1988,Umrigar_diffusionMonteCarlo_1993,PhysRevA.97.032307,PhysRevB.105.235144,PhysRevB.103.155135}. 

In this paper, we use a trial (or guiding, for DMC) wavefunction of the standard 
Jastrow-Slater form $\Psi_T({\bf R_{\uparrow}},{\bf R_{\downarrow}})=~J({\bf 
R_{\uparrow}},{\bf R_{\downarrow}})D_\uparrow({\bf R_\uparrow})D_\downarrow({\bf 
R_\downarrow})$, where ${\bf R_{\sigma}}$ refers to all the coordinates of 
fermions with spin $\sigma$. The antisymmetry of the wavefunction upon exchange 
of two fermions with the same spin is guaranteed by the Slater determinants 
$D_{\sigma}$, where the single-particle orbitals are taken to be plane waves 
$e^{i {\bf k}_l {\bf r}_i}$. In order to fulfill PBC, the allowed ${\bf k}_l$ 
have components $(k_x,k_y,k_z)=(n_x,n_y,n_z)2\pi/L$, where $n_{x,y,z}$ are 
integer numbers. The Jastrow factor is a symmetric product of two-body 
correlations $J({\bf R_{\uparrow}},{\bf 
R_{\downarrow}})=\prod_{i,j}^{N_\uparrow,N_\downarrow}f_{\uparrow\downarrow}(r_{
i j})\prod_{i,i^\prime}^{N_\uparrow}f_{\uparrow\uparrow}(r_{i 
i^\prime})\prod_{j,j^\prime}^{N_\downarrow}f_{\downarrow\downarrow}(r_{j 
j^\prime})$. Notice that the pair correlations $f_{\sigma\sigma^\prime}(r)$ 
depend only on relative distances and that we take them to be the analytical 
$s$-wave solution of the corresponding two-body problem, even in the same-spin 
case $\sigma^\prime=\sigma$. The reason for this choice is that the purpose of 
the Jastrow factor is to smoothen the local energy $E_L({\bf R_{\uparrow}},{\bf 
R_{\downarrow}})=\langle{\bf R_{\uparrow}},{\bf 
R_{\downarrow}}|\hat{H}|\Psi_T\rangle/\langle{\bf R_{\uparrow}},{\bf 
R_{\downarrow}}|\Psi_T\rangle$, while proper antisymmetry is taken care of by 
$D_{\sigma}$. This is arguably an accurate choice for moderate interactions, 
while for strongly repulsive or attractive interactions, a Pfaffian wavefunction 
could be more suited, where two-particle orbitals are directly antisymmetrized 
and can themselves be 
antisymmetric~\cite{Bajdich_PfaffianPairingWave_2006,
Bajdich_Pfaffianpairingbackflow_2008}. 

In determining the pair correlations $f_{\sigma\sigma^\prime}(r)$, we impose the boundary conditions $f_{\sigma\sigma^\prime}(\bar{R}_{\sigma\sigma^\prime})=1$ and $f^\prime_{\sigma\sigma^\prime}(\bar{R}_{\sigma\sigma^\prime})=0$, where $\bar{R}_{\sigma\sigma^\prime}$ are variational parameters that play the role of a healing length. We optimize $\bar{R}_{\sigma\sigma^\prime}$ with the stochastic reconfiguration method~\cite{CasulaCorrelatedgeminalwave2004}, which requires the evaluation of the analytic derivatives of the trial wavefunction with respect to the parameters. Since the above boundary conditions yield implicit equations for $\bar{R}_{\sigma\sigma^\prime}$, we use Dini's theorem on implicit function derivation. 

\subsection{Dispersion relation for single particle excitations}\label{subsec:dispersion}
In the context of QMC simulations of fermionic systems, the evaluation of the Fermi-liquid theory effective mass $m^*$ is usually performed via the calculation of the single-particle dispersion through energy estimates using different nodal surfaces, compatible with finite total momentum (see Ref.~\cite{Azadi_QuasiparticleEffectiveMass_2021} for a recent account and results on the electron gas).

For a fully polarized system, let $E(N)$ be the total energy of $N_\uparrow=N$ 
fermions with density $n$, where $N$ corresponds to filling only closed shells 
(we consider $N=33, 57, 81, 123, 171$) and the total momentum is zero, since all 
the single-particle orbitals in the Slater determinant are matched and maximally 
symmetric. Each closed shell contains all the wavevectors corresponding to the 
same integer modulus square $M=n_x^2+n_y^2+n_z^2$. The last closed shell in the 
reference configuration defines the Fermi wavevector for this finite system, 
$k_F^{N}=2\pi\sqrt{M_{\text{max}}}/L$. By keeping the volume $V=N/n$ fixed, we 
add a fermion with wavevector ${\bf k}$ in a shell with $M>M_{\text{max}}$, 
corresponding to momentum $\hbar k=\hbar|${\bf k}$|=2\pi\hbar\sqrt{M}/L$, and we 
denote the resulting total energy by $E(N+1,{\bf k})$. Alternatively, we add a 
vacancy, by removing a fermion with $M\le M_{\text{max}}$, and we denote the 
resulting total energy by $E(N-1,{\bf k})$. For a homogeneous system, the choice 
of the specific wavevector within a shell is irrelevant. The dispersion relation 
is defined as
\begin{equation}\label{eq:dispersion}
    \varepsilon({\bf k})+\mu = \begin{cases}
    E(N+1,{\bf k})-E(N)& k>k_F^{N}\\ 
    E(N)-E(N-1,{\bf k})& k\le k_F^{N}
    \end{cases}
\end{equation}
where $\mu$ is the chemical potential of the $\uparrow$ particles. By definition of $\mu$, $\varepsilon({\bf k})=0$ for momenta in the last reference shell, when $|{\bf k}|=k_F^N$. We are interested in the slope of the dispersion relation close to the Fermi momentum because the definition of the effective mass is 
\begin{equation}\label{eq:fiteffectivemass}
   \varepsilon({\bf k}) \underset{k\sim k_F^N}{\simeq} \frac{\hbar^2 k_F^N}{m^*}(k-k_F^N) = 2 \frac{m}{m^*} E_F^N (\tilde{k}-1)
\end{equation}
where $E_F^N=\hbar^2 (k_F^N)^2/2m$ is the Fermi energy for the finite system of $N$ fermions, and $\tilde{k}=k/k_F^N$. We fit Eq.~\eqref{eq:fiteffectivemass}, with the inclusion of a constant offset (namely the chemical potential) and a quadratic term $\propto(\tilde{k}-1)^2$, to the QMC data of Eq.~\eqref{eq:dispersion} for various wavevectors. The range of validity of the quadratic expansion around the Fermi surface could in principle be narrow; for a finite system, the discrete grid of wavevectors puts a lower bound on the size of the momentum range that can be fitted with sufficient precision, therefore finite-size effects are introduced and different numbers of fermions have to be considered, in order to assess the thermodynamic limit. 

\subsection{Finite-size effects and effective mass}\label{subsec:finitemass}
The dispersion approach requires a large number of simulations, involving both different particle numbers and different momenta. An alternative approach was proposed, that involves the evaluation of only the zero-total-momentum ground-state energy for different particle numbers~\cite{Tanatar_Groundstatetwodimensional_1989,Kwon_QuantumMonteCarlo_1994}. In spite of its relatively limited computational cost, this method has been much less frequently used, perhaps due to its reliance on the validity of Fermi-liquid theory and its rigorous demonstrability only in the weakly interacting regime. We report here its derivation for completeness.

Fermi-liquid theory~\cite{landau1957theory,Abrikosov_theoryfermiliquid_1959} postulates a one-by-one correspondence between particles of a noninteracting Fermi gas and quasiparticles in the Fermi liquid, which are therefore labeled by wavevector ${\bf k}$, in a homogeneous system (here we neglect spin since we are considering a fully polarized system). Quasiparticle energies $\varepsilon$ are determined in a self-consistent fashion as the functional derivative of the energy density $e$ with respect to the variation of their distributions $n({\bf k})$ upon perturbation:
\begin{equation}
    \delta e = \int d\tau \varepsilon({\bf k})\delta n({\bf k})\,,
\end{equation}
with $d\tau=d{\bf k}/(2\pi)^d$, where $d$ is the dimensionality. In general, the quasiparticle energies depend on the variation of the occupations, and a linear expansion is considered $\varepsilon({\bf k})=\varepsilon_0({\bf k})+\int d\tau^\prime f({\bf k},{\bf k}^\prime)\delta n({\bf k}^\prime)$. For the moment we neglect the quasiparticle interaction function $f$, and only consider the equilibrium term $\varepsilon_0({\bf k})$, which can be expanded close to the Fermi surface as 
in Eq.~\eqref{eq:fiteffectivemass}. 

We now specialize our discussion to a finite system of $N$ fermions in a cubic box of volume $V$ with PBC at zero temperature. The equilibrium distribution of quasiparticles in this case is a sum over the wavevectors $\mathcal{K}=\{{\bf k}_i\}$ allowed by PBC, up to the Fermi wavevector: $n_0({\bf k})=(2\pi)^d \sum_{{\bf k}_i\in\mathcal{K}}\delta({\bf k}-{\bf k}_i)/V$, with the normalization constraint on the total density $n = \int d\tau n_0({\bf k})$. We consider particle numbers that correspond to filling only closed shells, such that the set $\mathcal{K}$ is closed upon all the system symmetries~\cite{Lin_Twistaveragedboundaryconditions_2001}. As a perturbation from the equilibrium of $N$ particles in a volume $V=N/n$, we consider a variation of number and volume such that the density remains fixed $n=N^\prime/V^\prime$~\cite{Kwon_QuantumMonteCarlo_1994}. The new equilibrium quasiparticle distribution is thus $n_0^\prime({\bf k})=(2\pi)^d \sum_{{\bf k}_i^\prime\in\mathcal{K}^\prime}\delta({\bf k}-{\bf k}_i^\prime)/V^\prime$, where the wavevectors in $\mathcal{K}^\prime$ are those compatible with PBC in volume $V^\prime$, and the same normalization as for $n_0({\bf k})$ holds. Let us consider the variation of the occupations $\delta n({\bf k})=n_0^\prime({\bf k})-n_0({\bf k})$: for sufficiently large $N$ and $N^\prime$, we can substitute $\delta n({\bf k})$ with its average in a neighborhood of ${\bf k}$ of size larger than the typical spacing between wavevector coordinates in $\mathcal{K}$ and $\mathcal{K}^\prime$, namely, $2\pi/V^{1/d}$ and $2\pi/{V^\prime}^{1/d}$, respectively. The resulting smeared function is close to zero for $k\ll k_F^{N},k_F^{N^\prime}$, since well inside the Fermi sphere the PBC wavevectors are equally spaced; obviously, it is zero for $k\gg k_F^{N},k_F^{N^\prime}$, while it can have large fluctuations for $k\simeq k_F^{N} \simeq k_F^{N^\prime}$, due to the different symmetry of the last closed shells for different particle numbers. For this reason, the difference in energy density between two Fermi liquids with different particle numbers is determined by the quasiparticles close to the Fermi surface, and we can safely extend Eq.~\eqref{eq:fiteffectivemass} to wavevectors far from the Fermi surface, where $\delta n\approx 0$. In this derivation, we can now assume that the dependence of $\mu$ and $k_F$ on the number of particles is negligible because it contributes only to second order in $\delta n$. It is convenient then to write $\varepsilon_0({\bf k})\approx \frac{\hbar^2}{2m^*} (k^2-k_F^2)$, so that 
\begin{multline}\label{eq:finitesize}
   \delta e \approx \int d\tau \left[\frac{\hbar^2}{2m^*} (k^2-k_F^2)\right]\delta n({\bf k}) \\
   =\frac{m}{m^*}\int d\tau \frac{\hbar^2k^2}{2m} \left(n_0^\prime({\bf k})-n_0({\bf k})\right) = \frac{m}{m^*} \left(t_0^\prime-t_0\right)\,,
\end{multline}
where we have used the normalization constraint of the occupations and the definition of the kinetic-energy density of the ideal Fermi gas $t_0=\int d\tau n_0({\bf k})\hbar^2k^2/2m$. 

Since $t_0$ is tabulated, relation \eqref{eq:finitesize} provides a relatively straightforward method to determine the effective mass with only the evaluation of the total energy of closed shell systems with different particle numbers. It also provides a finite-size correction to the energies from QMC simulations of $N$ fermions, by letting $N^\prime\to\infty$ and using $t_0^\infty=3E_F/5$ for a polarized 3D system, with straightforward generalization to other polarizations. We use this correction when reporting energies in Secs.~\ref{sec:polarized} and \ref{sec:balanced}, but approximating $m^*=m$. The errorbars then include QMC statistical uncertainty and an estimate of residual finite-size effects based on the above correction. These are summed in quadrature and the result is smaller than the symbol size.

In order to take into account also second order corrections in the variation of occupations and the role of the quasiparticle interaction $f$, Eq.~\eqref{eq:finitesize} can be extended with a phenomenological term of type $c/N^\nu$, where $c$ and $\nu$ are parameters to be fitted from the QMC data~\cite{Tanatar_Groundstatetwodimensional_1989,Drummond_Finitesizeerrorscontinuum_2008}. This correction is expected to be relevant only for small number of fermions, or when the interactions are strong. 

\section{Analysis of the fully polarized Fermi gas}\label{sec:polarized}
\begin{figure}[!tb]
    \centering
    \includegraphics[width=0.95\columnwidth]{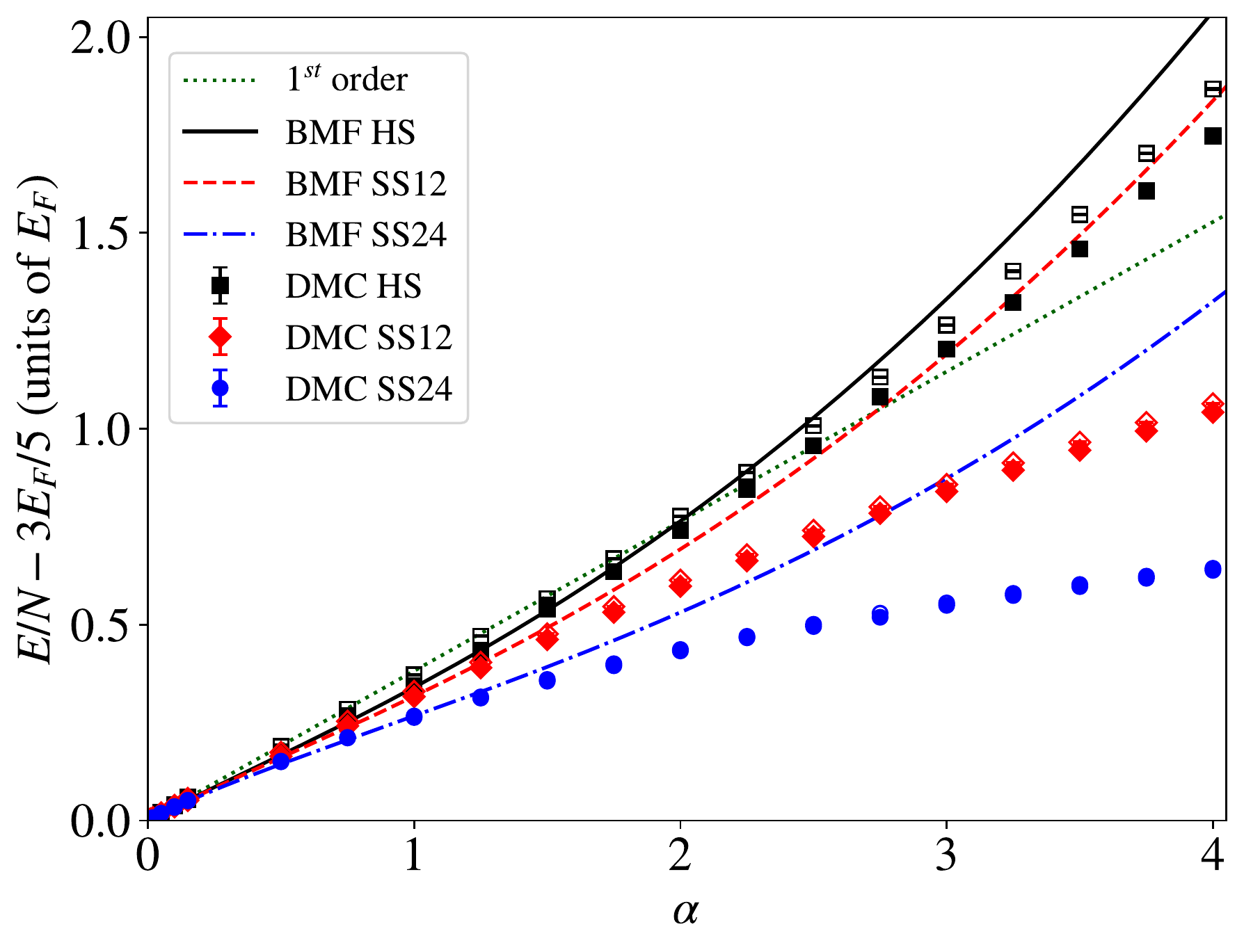}
    \caption{DMC interaction energy of a fully polarized Fermi gas interacting with the HS, SS12, and SS24 potentials as a function of $\alpha$, compared to the linear and BMF predictions. Empty symbols correspond to VMC simulations.}
    \label{fig:eqstateUU}
\end{figure}
The main focus of this paper is on a spin-polarized Fermi gas, where $n=n_\uparrow$ and interaction effects at low energy are mainly accounted for by $p$-wave contributions.

The zero-temperature Fermi-liquid theory treatment of such a system has been recently perturbatively studied in Ref.~\cite{Ding_FermiLiquidDescriptionSingleComponent_2019}. In particular, the correction $\epsilon$ to the non-interacting energy per particle, $E/N = E_F \left(3/5+\epsilon \right)$, has been calculated to be, to the second order in the scattering volume and the dominant order in the effective range,

\begin{equation}
\label{eqsinglecomp}
\epsilon = \frac{6\alpha}{5\pi}-\left(\frac{9}{35\pi \zeta}-\frac{2066-312\log{2}}{1155\pi^2}\right)\alpha^2\,,
\end{equation}
where $\alpha=k_F^3 v$ and $\zeta=k_F R$. Both $\alpha$ and $\zeta$ are small in the weakly interacting limit (see App.~\ref{app:scattering}). Strikingly, the effective range appears in the denominator, indicating that the coefficient of the second-order term in $v$ is strongly dependent on the effective range, in the weakly interacting regime. 

The above expression has been obtained for a generic potential; however, when considering a specific class of potentials, like the SS ones and their limit, the HS one, $\zeta$ and $\alpha$ are not independent: in this case Eqs.\eqref{eq:appv},\eqref{eq:appR} allow one to consider $\zeta$ as an implicit function of $\alpha$, or equivalently of the density $n R_S^3$, having fixed the dimensionless strength $K_0$ to three representative values. Therefore, Eq.~\eqref{eqsinglecomp} can be viewed as an expansion in $\alpha$, where the first and the last terms are the effective-range-independent linear and quadratic contributions, respectively, while the second term is of order $\alpha^{5/3}$. The coefficient of this term is very large for small $K_0$, while it decreases when approaching the HS limit. As a result of the competition between the $\alpha^{5/3}$ and $\alpha^2$ terms, the BMF correction to the energy is negative for small $\alpha$, but changes sign for sufficiently large $\alpha$.

In Fig.~\ref{fig:eqstateUU}, we compare the above perturbative prediction for the zero-temperature equation of state with VMC (empty symbols) and DMC (full symbols) simulation results for the HS, SS12, and SS24 potentials, with varying $\alpha\le 4$. The same $\alpha$ corresponds to different $\zeta$, depending on the potential. VMC and DMC results show very few discrepancies, which indicates that the variational optimization of the Jastrow factor is satisfactory. All the considered systems follow the perturbative effective-range-independent linear prediction for $\alpha\lesssim 0.5$, while BMF effective-range-dependent contributions are manifest for stronger repulsion. Consistently with Eq.~\eqref{eqsinglecomp}, departures from the linear behavior are larger the softer the repulsive potential is (corresponding to smaller effective-range). Also, the regime of quantitative agreement between perturbative and DMC predictions goes from $\alpha\lesssim 1$ for the SS24 potential to $\alpha\lesssim 2$ for the HS potential. In the latter case one can also observe a change of concavity of the equation of state, in qualitative agreement with the discussion in the previous paragraph.

Ref.~\cite{Ding_FermiLiquidDescriptionSingleComponent_2019} also provides the BMF expression for the effective mass of the quasiparticles in the fully polarized case, which has been deduced to be

\begin{equation}\label{eq:perturbeffectivemass}
    \frac{m}{m^*} = 1+\frac{2\alpha}{\pi}-\left[\frac{1}{\pi \zeta}-\frac{8(313-426\log{2})}{315\pi^2}\right]\alpha^2\,.
\end{equation}
Differently from the $s$-wave case, this expression varies from the non-interacting case $m^*=m$ already at the mean-field level. In the range of $\alpha$ that we are considering, the BMF correction is dominated by the negative effective-range-dependent contribution, since the last term has a very small coefficient.

\begin{figure}[!tb]
    \centering
    \includegraphics[width=0.95\columnwidth]{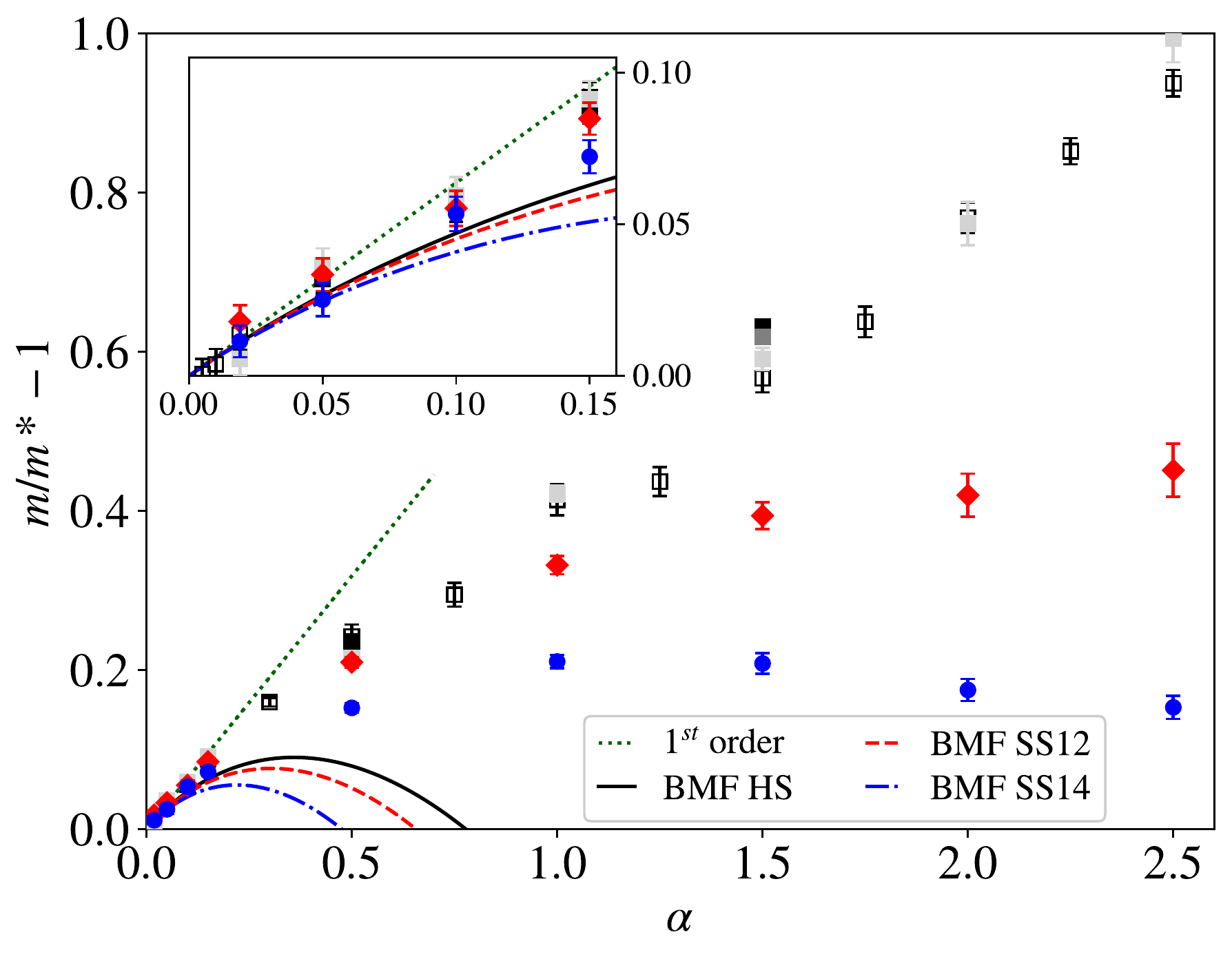}
    \caption{VMC determination of the inverse effective mass, offset by $1$ as a function of $\alpha$, compared to the perturbative prediction of Eq.\eqref{eq:perturbeffectivemass}. Shown are results for both the finite-size method (empty squares, for the HS model) and the dispersion method with $N=123$ particles (light gray filled squares, red diamonds, and blue circles for the HS, SS12, and SS24 models, respectively). For selected values of $\alpha$, the dispersion method for the HS model is also used with $N=33$ (black filled squares) and $N=57$ (dark gray squares). Inset: the range $\alpha\leq 0.15$ is highlighted.}
    \label{fig:mass}
\end{figure}

We show our QMC results for the inverse effective mass in Fig.~\ref{fig:mass}. Due to the demanding computational requirements, for the effective mass determination we only employ VMC simulations, whose energies are very close to the DMC ones (see Fig.~\ref{fig:eqstateUU}). For some points, we have in fact evaluated the effective mass also with DMC (not shown), finding compatible values with VMC, albeit with larger uncertainty bars. We consider both the fit of the single-particle dispersion of Sec.~\ref{subsec:dispersion}, for $N=N_\uparrow=123$ fermions (and also for $N=33,57$ with the HS model at some selected values of $\alpha$), and with the finite-size method of Sec.~\ref{subsec:finitemass} fitting results for $N=33, 57, 81, 123$, and $N=171$ in some cases. For both methods, we increase the uncertainty bars up to $0.02$ when we notice that removing some particle sizes or momenta from the fits yields differences in the output effective mass exceeding the error bars. We also notice that for $\alpha\gtrsim 1$, a correction of type $1/N^2$, as mentioned in Sec.~\ref{subsec:finitemass}, is important to obtain reduced $\chi^2\simeq 1$, without significantly affecting the value of $m^*$. 

We find that the two ways of evaluating the inverse effective mass in QMC are consistent, provided that the dispersion relation is calculated for a sufficiently high number of fermions. In particular, we observe that the HS results at $\alpha=0.5$ and $1.5$ calculated with the dispersion method with $N=33, 57, 123$ particles are consistent with each other and with the result from the finite-size method within $\simeq0.05$. We find that the QMC results are consistent with the effective-range-independent linear correction predicted by Eq.~\eqref{eq:perturbeffectivemass} for $\alpha<0.1$. A strong dependency on the chosen potential, namely, on the effective range, is observed for $\alpha\simeq 0.15$. Also, in this regime, the QMC results with different potentials depart from the BMF correction in Eq.~\eqref{eq:perturbeffectivemass}. This is in striking contrast with the consistency of the perturbative and QMC methods for the equation of state, which is generically apparent up to $\alpha\simeq 1$. For $\alpha\lesssim 0.1$, the errorbars are too large to make conclusive statements regarding the accuracy of the BMF correction. However, further reducing QMC errorbars in this regime would require one not only to increase statistics, but also to better assess the possible residual role of finite-size effects and of the chosen fitting function, significantly increasing the computational burden.

The origin of the discrepancy of beyond-mean-field $p$-wave effects in the effective mass can be two-fold: on the perturbative side, it might be that convergence in this quantity requires resummation of higher order terms, especially for small effective range; on the QMC side, it might be that the nodal surface in both the VMC and DMC calculations plays an important role, as for the electron gas~\cite{Azadi_QuasiparticleEffectiveMass_2021}, a question that will be worth investigating in the future.

\section{Role of equal-spin interaction in the spin-balanced case}\label{sec:balanced}

\begin{figure}[!tb]
    \centering
    \includegraphics[width=0.95\columnwidth]{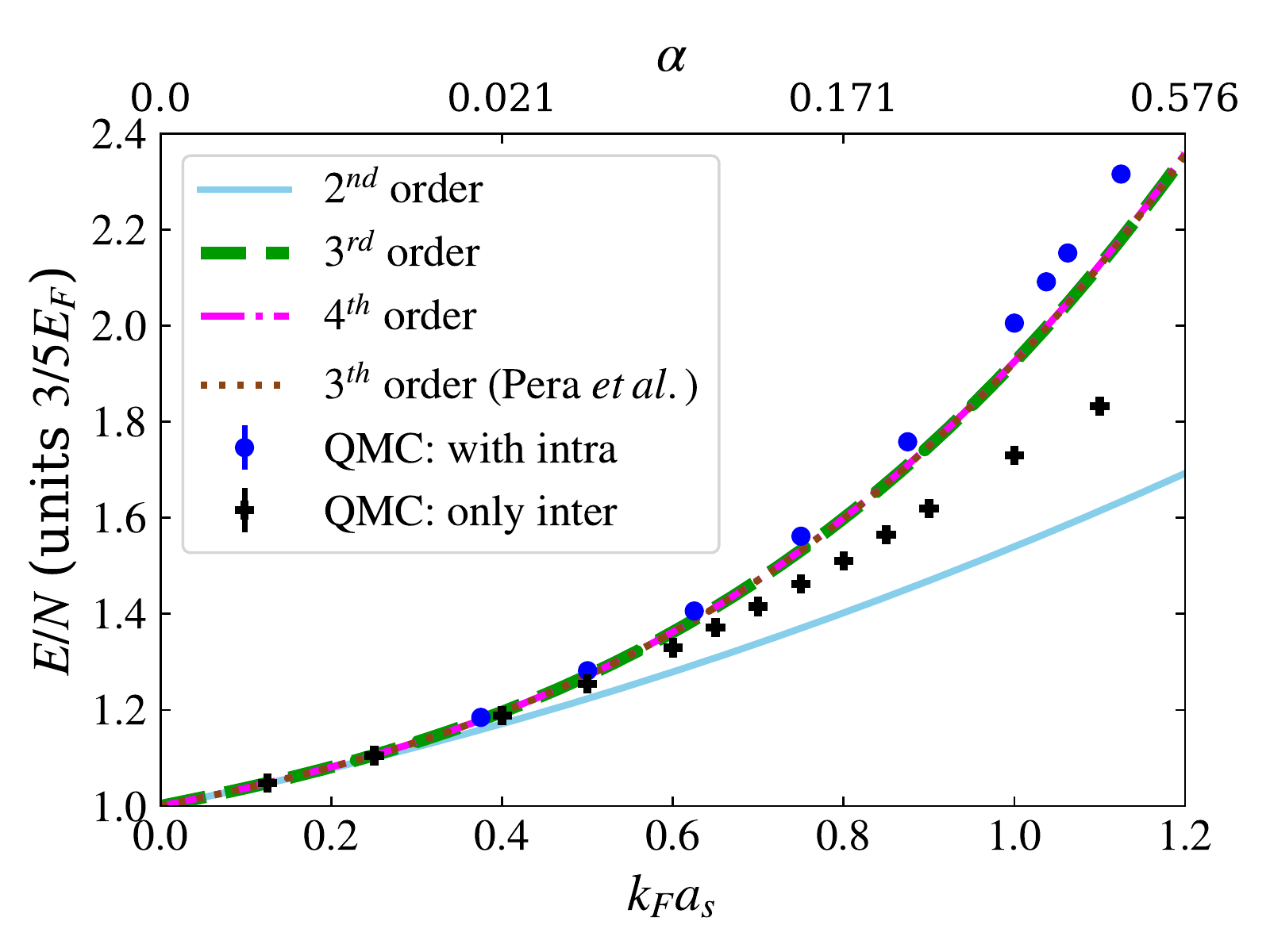}
    \caption{DMC equation of state (blue circles) of a spin-balanced Fermi gas 
with hard-sphere interaction both in the opposite and in the equal spin sectors. 
The energy per particle $E/N$ is plotted vs the interaction parameter $k_F 
a_s$. The upper horizontal axis reports the corresponding values of $\alpha$. 
This is compared to perturbative expansions with increasing order 
from~\cite{Wellenhofer_Effectivefieldtheory_2021}, and with the third-order 
expansion by Pera \emph{et 
al.}~\cite{https://doi.org/10.48550/arxiv.2206.06932}. As a reference, the DMC 
results in the absence of intraspecies interactions are also shown (black 
crosses)~\cite{2010Pilati,ma2012density}.}
    \label{fig:eqstate-balanced}
\end{figure}

This section focuses on balanced (i.e., unpolarized) Fermi gases, namely, on the case $n_\uparrow=n_\downarrow=n/2$. 
In this case, the Fermi wavevector and energy are $k_F=(3\pi^2n)^{1/3}$ and $E_F=\hbar^2k_F^2/2m$, respectively.
The HS potential is adopted to model both intra- and interspecies scatterings. 
It is worth emphasizing that, with this model, all scattering parameters, including the s-wave scattering length and effective range, as well as the p-wave scattering volume and effective range, are fixed by the HS diameter $R_S$ only. In particular, the s-wave scattering length is $a_s=R_S$, and the interaction parameter considered in Section~\ref{sec:polarized} can be written as $\alpha=\left(3\pi^2n\right)^3\nu=\left(k_F R_S\right)^3/3$. Following the standard convention, here we discuss our results in terms of the interaction parameter $k_F a_s$.
Our fixed-node DMC results are compared with various perturbative expansions from the literature. 
One of our main goals is to inspect the role of intraspecies interactions, making comparison with the case where only interspecies interactions are accounted for in the DMC simulations. 
Notice that intraspecies interactions affect the itinerant ferromagnetic transition~\cite{AriasdeSaavedra_Ferromagnetictransitiontwocomponent_2012,PhysRevA.93.051605,PhysRevA.102.053301}.
The DMC energy per particle $E/N$ is shown in Fig.~\ref{fig:eqstate-balanced}, as a function of $k_F a_s$. 
In the broad regime $k_Fa_s\lesssim 0.5$, the results obtained with only interspecies interactions closely match the corresponding results including also the intraspecies channels. However, for $k_Fa_s \simeq 1$, sizable deviations occur. This indicates the increased role of $p$-wave scattering in this regime.
Notably, the perturbative equation of state for this system has been recently extended up to the fourth order~\cite{Wellenhofer_Effectivefieldtheory_2021}. When specialized to the HS model, the equation of state reads
\begin{equation}
\frac{E}{N}=\frac{3}{5}E_F \sum_{j=0}^{j_{\mathrm{max}}} C_j\left(k_F a_s\right)^j,
\end{equation}
where $(C_0,C_1,C_2,C_3,C_4)=(1,0.354,0.186,0.384,0.001)$, and the integer $j_{\mathrm{max}} \leq 4$ is the chosen expansion order in the interaction parameter $k_F a_s$.
A third-order expansion has been provided also in Ref.\cite{https://doi.org/10.48550/arxiv.2206.06932} for arbitrary number of components and populations imbalances. When specialized to two balanced components, this expansion closely agrees with the corresponding result of Ref.~\cite{Wellenhofer_Effectivefieldtheory_2021} with $j_{\mathrm{max}} =3$ (see Fig.~\ref{fig:eqstate-balanced}). 
It is worth pointing out that, when specialized to the single-component case, the third-order expansion of Ref.\cite{https://doi.org/10.48550/arxiv.2206.06932} corresponds to Eq.~\eqref{eqsinglecomp} truncated to the first order in the scattering volume $v$. In particular, effects due to $p$-wave effective range $R$ are not accounted for.
From Fig.~\ref{fig:eqstate-balanced}, one also notices that the contribution of the fourth-order term ($j=4$) is very small.
Notice that, also in the balanced case, these third- and fourth-order expansions do not account for the effect of the $p$-wave effective range $R$.
Notably, these expansions barely deviate from the DMC results including also intraspecies scattering. The residual discrepancy might be attributed to the fixed-node constraint, meaning that a more accurate nodal surface or an unbiased released node technique would provide lower energies, in agreement with the perturbation theory. On the other hand, it is also plausible that beyond-fourth-order terms, or terms depending on $R$, would provide the missing contribution to exactly match the DMC results.

\section{Conclusion}\label{sec:conclusions}
The role of intraspecies interactions in (pseudo-)spin-1/2 Fermi gases with short-range repulsive potentials has been investigated via fixed-node diffusion and variational Monte Carlo simulations. The ground-state energy per particle for the fully polarized gas has been determined and the roles of the $p$-wave scattering volume and of the $p$-wave effective range have been separately elucidated. The effective mass has also been determined, finding deviations from the recently derived perturbative expansion already for moderately strong interaction strength, where agreement is instead found for the equation of state.
Furthermore, the equation of state for the balanced configuration has been determined, making comparison with previous studies that considered only interspecies scattering. Notably, we have found good agreement with the recently derived fourth-order perturbative expansion.

We hope that our findings will favor the further development of many-body 
theories for Fermi gases, which should take into account $p$-wave scattering 
effects~\cite{ALARCON2022168741,Alarcon_Ultracoldspinbalancedfermionic_2022,
Beane_precisionFermiliquidtheory_2023}. Indeed, ultracold atom experiments 
reached high-density regimes where these effects become relevant.
The $p$-wave parameters, scattering volume and effective range, are also relevant to describe $p$-wave Feshbach resonances in the resonant regime~\cite{Zhang_Scatteringamplitudeultracold_2010,Luciuk_Evidenceuniversalrelations_2016,Ahmed-Braun_Probingopenclosedchannel_2021}. Future research should check whether the regime of reliability of the perturbative expansion is reduced or not, when considering different potentials with the same scattering volume and effective range, establishing universality in terms of these two scattering parameters. It will also be relevant to study the effective mass by means of more computationally demanding, but more accurate, DMC simulations, possibly including backflow correlations~\cite{Motta_Implementationlinearmethod_2015}.

Strictly speaking, the results presented in this paper and in 
\cite{Ding_FermiLiquidDescriptionSingleComponent_2019} are valid for short-range 
potentials, which admit an effective-range expansion in powers of $k^2$ for the 
scattering amplitude of any partial wave. This is also the case for the $s$-wave 
amplitude of realistic atomic potentials asymptotically decaying with a van der 
Waals tail $-C_6/r^6$. However, scattering theory for such potentials predicts 
that the effective-range expansion of the $p$-wave amplitude contains also a 
term proportional to $k$ \cite{Gao_Quantumdefecttheoryatomic_1998}. On the one 
hand, it has been argued that this term is negligible when considering large 
$p$-wave scattering volumes, for example in two-channel models describing 
$p$-wave Feshbach resonances \cite{Zhang_Scatteringamplitudeultracold_2010}; on 
the other hand, this fact highlights the opportunity to investigate more the 
many-body consequences of a van der Waals tail in the weakly interacting regime 
of polarized Fermi gases, and we leave this endeavor for future studies. Van der 
Waals potentials also fulfill universal relations between scattering parameters 
of different partial waves \cite{Idziaszek10}. This property, together with 
density shifts in an optical clock, was used in 
Refs.~\cite{Martin_QuantumManyBodySpin_2013,
Zhang_SpectroscopicobservationSU_2014} to estimate the values 
$v_{gg}^{1/3}=76.6(4)a_0$, $v_{ee}^{1/3}=-119(18)a_0$ and 
$v_{ge+}^{1/3}=-169(23)a_0$, for two $^{87}Sr$ atoms in a fully polarized 
nuclear spin configuration and symmetric electronic states $|gg\rangle$, 
$|ee\rangle$ and $(|ge\rangle+|eg\rangle)/\sqrt{2}$, respectively, where the 
relevant (pseudo-)spin levels are $g=^1$S$_0$ and $e=^3$P$_0$. These scattering 
parameters correspond to $\alpha\ll 1$ at the densities of standard atomic 
clocks, indicating that the linear contribution due to the $p$-wave scattering 
volume is sufficient to estimate density shifts in such cases. A relevant 
extension of this paper is to consider different confinement geometries, such as 
the quasi-two-dimensional ones, that allow one to reach the Lamb-Dicke regime in 
optical lattice clocks. Finally, combining accurate QMC methods with the 
simulation of laser-driven interacting atomic systems with generic polarization 
and confinement is still an open problem.

Data to reproduce the figures in this paper are available online (Ref.~\cite{gianluca_bertaina_2023_7516667}).

\acknowledgments
We acknowledge useful comments by S. Zhang.
G.B. acknowledges useful discussions with P. Pieri, N. Cuzzuol and J. D'Alberto, concerning the calculation of effective masses. 
S.P. acknowledges interesting discussions with J. Pera and J. Boronat.
This work was partially supported by the Italian Ministry of University and Research under the PRIN2017 project CEnTraL (Grant No. 20172H2SC4). S.P. acknowledges PRACE for awarding access to the Fenix Infrastructure resources at Cineca, which are partially funded by the European Union’s Horizon 2020 research and innovation program through the ICEI project under Grant No. 800858. G.B. acknowledges the CINECA awards IscrC-SEMIPRO (2019) and IscrC-BF2D (2021), for the availability of high performance computing resources and support.

\appendix \section{Partial wave scattering amplitude in three dimensions}\label{app:scattering}

\begin{figure}[!tbp]
    \centering
    \includegraphics[width=0.95\columnwidth]{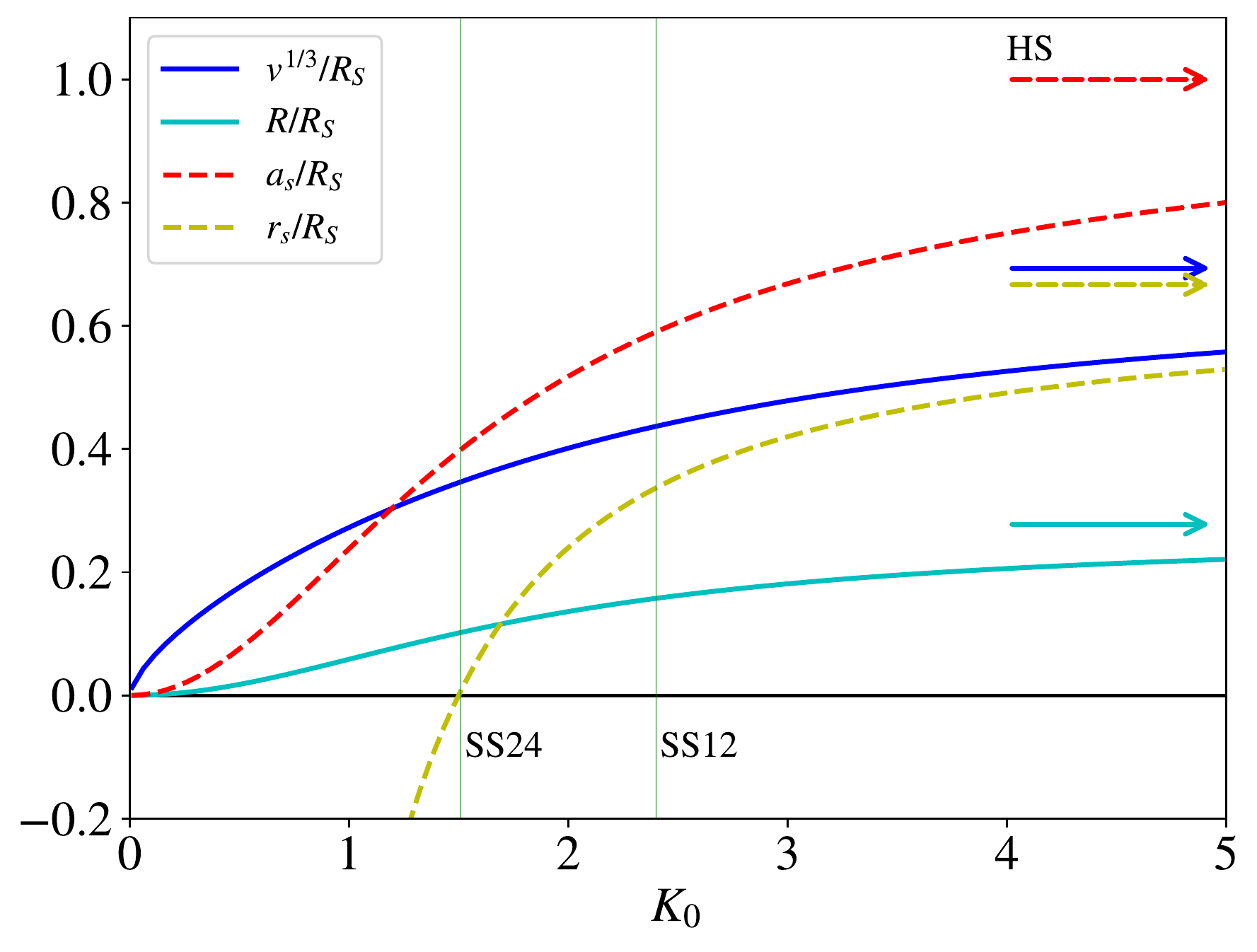}
    \caption{Low-energy scattering parameters of the SS potential, in units of the range $R_S$, as a function of the dimensionless strength $K_0$. Two vertical lines mark the two considered SS potentials, while arrows indicate the HS limits.}
    \label{fig:scatteringSS}
\end{figure}

The expansion of the low-energy scattering amplitude in three dimensions, for angular momentum number $l$, reads:
\begin{equation}
    f_l(k)^{-1}= k \cot\left[\delta_l(k)\right] - i k = k^{-2l}\left(- \frac{1}{a_l} + \frac{r_l}{2} k^2 + O[k^4]\right) - i k\,,
\end{equation}
where the relative momentum is $k^2=2\mu E/\hbar^2$, $E$ is the kinetic energy of the two scattering particles and $\mu=m/2$ is the reduced mass for equal mass particles. In the $s$-wave case $f_0(k)^{-1}= \left(- \frac{1}{a_s} + \frac{r_s}{2} k^2 + O[k^4]\right) - i k$ where $a_s$ and $r_s$ are the $s$-wave scattering length and effective range, respectively.

The $p$-wave case reads
\begin{equation}
    f_1(k)^{-1}= \left(- \frac{1}{v k^2} - \frac{1}{2 R} + O[k^2]\right) - i k\,,
\end{equation}
where, for dimensionality reasons, $v$ is the $p$-wave scattering volume, while $R$ is the $p$-wave effective range. Notice that some authors define the $p$-wave effective range as the inverse of $R$, which is then an effective momentum, possibly including the factor $2$ in the definition~\cite{Hammer_Causalityeffectiverange_2010}. Notice also the minus sign, that we introduced to follow the convention used in ~\cite{Ding_FermiLiquidDescriptionSingleComponent_2019}.

Here, we consider the SS model potential and adapt the results in~\cite{Hammer_Causalityeffectiverange_2010}, with $K_0^2 = m V_{0} R_{S}^2/\hbar^2$.
The $p$-wave scattering volume is:
\begin{equation}\label{eq:appv}
v =\frac{R_{S}^3}{3}\left(1-3\frac{K_0 C_0 - 1}{K_0^2}\right)\,,
\end{equation}
with $C_0=\coth{K_0}$. The $p$-wave effective range is:
\begin{equation}\label{eq:appR}
R=\frac{5 R_S}{9} \frac{\left(K_0^2+3-3 K_0 C_0\right)^2}{
15K_0^2 C_0^2-5\left(2K_0^2+3\right) K_0 C_0+\left(2K_0^2+5\right)K_0^2}.
\end{equation}

In the HS limit when $V_0\to \infty$, one obtains $v=R_S^3/3$ and $R=5 R_S/18$. Some authors therefore also use the auxiliary definition for a $p$-wave scattering length $a_p=(3v)^{1/3}$, because it results in $a_p=R_S$ for the HS potential. For $V_0\to 0$, one obtains $v \to R_S^3 K_0^2/45$ and $R \to 7 R_S K_0^2/90$. So, in the weakly interacting limit both the scattering volume and the effective range are positive (using the convention in Ref.~\cite{Ding_FermiLiquidDescriptionSingleComponent_2019}) and small. 

For completeness we report here also the $s$-wave scattering length and effective range for the SS potential:
\begin{equation}
    a_s = R_{S}\left(1-1/C_0 K_0\right)\;,
\end{equation}
\begin{equation}
r_s =R_{S}\left(1-\frac{1}{3(1-1/C_0 K_0)^2}+\frac{1}{K_0^2(1-1/C_0 K_0)}\right)\;.
\end{equation}

In the HS case, one has $a_s=R_S$ and $r_s=2R_S/3$, while in the weakly interacting regime one obtains $a_s \to R_S K_0^2/3$ and $r_s \to -6R_S/5K_0^2$. Notice that $r_s$ changes sign and is infinite and negative for small $V_0$.

In Fig.~\ref{fig:scatteringSS}, we show the $p$-wave scattering volume and effective range for the SS potential, together with their $s$-wave counterparts. 

\bibliography{main}
\end{document}